\documentclass[pra,aps,twocolumn,psfig,epsfig]{revtex4}
\usepackage{color}
\usepackage{graphicx,amssymb,epsfig,epsf,amsmath}

\begin{document}

\title{Correlated many-body calculation to study characteristics of Shannon information entropy for ultracold trapped interacting bosons.}

\author{Sudip Kumar Haldar$^{1}$, Barnali Chakrabarti$^{2}$, Tapan Kumar Das$^{3}$, Anindya Biswas$^{4}$}

\affiliation{
$^{1}$Department of Physics, Lady Brabourne College, P 1/2 Suhrawardy Avenue, Kolkata 700017, India.\\
$^{2}$Department of physics, Kalyani University, Kalyani, Nadia, West Bengal, India, Pin: 741235.\\
$^{3}$Department of Physics, University of Calcutta, 92 A.P.C. Road, Kolkata-700009, India.\\
$^{4}$Harish-Chandra Research Institute, Chhatnag Road, Jhunsi, Allahabad-211019, India.}

\begin{abstract}
 A correlated many-body calculation is presented to characterize the Shannon information entropy of trapped interacting bosons. We reformulate the one-body Shannon information entropy in terms of the one-body probability density. The minimum limit of the entropy uncertainty relation (EUR) is approached by making $N$ very small in our numerical work. We examine the effect of correlations in the calculation of information entropy. Comparison with the mean-field result shows that the correlated basis function is indeed required to characterize the important features of the information entropies. We also accurately calculate the point of critical instability of an attractive BEC, which is in close agreement with the experimental value. Next we calculate two-body entropies in position and momentum spaces and study quantum correlations in the attractive BEC. 
\end{abstract}

\maketitle
\section{Introduction}
In recent years, the information theoretic methods play an interesting and important role in the study of quantum mechanical systems such as nuclei, atomic cluster, confined atoms etc~\cite{Ohya,S,E,C,CP,Ch}. The extraction of probability densities in both the position and momentum spaces and the associated calculation of information entropy provide some additional relevant information of the system. Recent experimental achievment of Bose Einstein condensation (BEC) in the dilute alkali atomic vapor~\cite{Anderson,Roberts,Cornish} opens a new avenue of research in this direction. The system of ultracold trapped atoms obey Bose Einstein statistics and a finite fraction of particles condense to the ground state below the critical temperature. The accumulated atoms in the single state can be described as a single quantum particle having nonlinear interaction. The presence of an external harmonic trapping potential allows for measurement both in the coordinate and momentum spaces. The interparticle interaction and the confining potential strongly influence the static and dynamic properties of the condensate. Correlations among the atoms in the external trap also play a key role even though the laboratory BEC is extremely dilute.

In this paper we present the results of numerical studies of Shannon information entropy of ultracold trapped interacting bosons in both the coordinate and momentum spaces.  We adopt two-body correlated basis function together with the realistic van der Waals potential for the description of the condensate at zero-temperature. From the condensate wave function we calculate the one-body density $R_{1}(\vec{r}_{k})$ and two-body density function $R_{2}(r_{ij})$, where $\vec{r}_{k}$ is the distance of $k$th  boson from the center of mass of the condensate and $r_{ij}$ is the relative separation of the $(ij)$ pair of bosons. The Shannon entropy in position space is defined as 
\begin{equation}
 S_r=-\int \rho(\vec{r}) \ln \rho(\vec{r}) d\vec{r} \hspace*{.1cm} 
\end{equation}
where $\rho(\vec{r})$ is either the one-body or the pair-density in the coordinate space. The momentum space Shannon entropy is described in similar way as 
\begin{equation}
 S_p = - \int \phi(\vec{p}) \ln \phi(\vec{p}) d\vec{p} \hspace*{.1cm}  
\end{equation}
where $\phi(\vec{p})$ represents the one-body or pair-momentum density. $\rho(\vec{r})$ and $\phi(\vec{p})$ are normalized to unity. Throughout this paper, we adopt the units, referred to as the oscillator units (o.u.), in which length and energy are expressed in units of $a_{ho}=\sqrt{\frac{\hbar}{m\omega}}$ and $\hbar\omega$ respectively, $\omega$ being trapping frequency. As densities are measured in this unit, all the entropy values presented here are also in o.u. The Shannon information entropies basically measure the uncertainty of the probability distribution in the respective spaces.
Using the position and momentum space entropies ($S_r$ and $S_p$ respectively), Bialynicki-Birula and Mycielski (BBM) derived a stronger version of the Heisenberg uncertainty relation~\cite{I}. For a three-dimensional system the total entropic sum has the form 
\begin{equation}
S=S_r+S_p \ge 3(1+\ \ln{ (\pi)}) \simeq 6.434  .
\end{equation}
This means that the conjugate position and momentum space information entropies $S_r$ and $S_p$ maintain an inverse relationship with each other. It signifies that when a system is strongly localized in position space, the corresponding information entropy and uncertainty in position space decreases. The corresponding momentum distribution becomes delocalized, therefore the information entropy and uncertainty in momentum space increases. It is important to note that BBM inequality~(Eq.(3)) presents a lower bound and the equality is maintained for a Gaussian wave function.

It is to be noted that information theory of correlated bosons in the external trap has been studied using uncorrelated mean-field theory~\cite{Massen}. The universal expression for the total entropy in different quantum systems takes the form $ S = a + b \ln N$, where $a$ and $b$ are the parameters that depend on the given system~\cite{S}. Although the choice of our system is quite similar to that of earlier studies, but the motivation of our present work is quite different and is as follows. First: as the interacting bosons in the external trap are correlated, the uncorrelated mean-field Gross-Pitaevskii (GP) equation may not reveal the correct information and characteristic features in the measures of Shannon entropy. In the GP equation, the interparticle interaction is taken as a contact interaction, whose strength is given by a single parameter $a_s$, the $s$-wave scattering length. However, the importance of a realistic, finite-range interaction has been pointed out~\cite{Gelt,Khan}. Hence the presence of a long attractive tail in a realistic interaction, like the van der Waal interaction will have important contributions to correlation function and will provide more realistic aspects. Thus our present calculation, keeping all-possible two-body correlations and realistic inter-atomic interaction will provide more accurate results than those obtained from GP and we may observe new features in the calculation of Shannon entropy and correlation properties of the confined bosons. Second: for the repulsive condensate $a_s$ is positive, therefore with increase in $a_s$ or increase in number of bosons $N$, the effective interaction parameter $Na_s$ increases. Thus for the repulsive condensate, with increase in $Na_s$ the system becomes less correlated, as the central density decreases. However for attaractive condensate $a_s<0$, the trend is in the opposite direction. It is well known that for attractive BEC, the condensate collapses when $N$ becomes equal to some critical number $N_{cr}$. Thus when $N$ is close to but less than $N_{cr}$, the condensate is highly correlated and for $N \geq N_{cr}$ the condensate collapses. This is a very crucial point where the study of information entropy may provide some new characteristic features of Shannon entropy and from that we may get very accurate value of the stability factor which is experimentally known~\cite{Roberts}. This is beyond the scope of study of mean-field GP equiation. We will also introduce correlation features to study the correlation properties of attractive condensate.

This paper is organised as follows. In Sec. II we briefly review our theoretical approach. In Sec. III we present our calculation and results of information entropy of the one-body density, entropy uncertainty relation (EUR) and the universal property of total entropy. This section also presents results of pair correlation and Shannon information entropy of two body density and statistical correlation function. Finally in Sec. IV, we draw our conclusions. 

\section{Methodology}
\subsection{Many-body calculation with correlated potential harmonic basis}

In this work we have employed our newly developed correlated potential harmonic expansion method (CPHEM) which we have succesfully used in our earlier studies of different properties of condensate~\cite{Sudip1,Pankaj1,Pankaj2,Anindya1,Anindya2}. Here we briefly describe the method. Details can be found in Ref.~\cite{Tapan,Das,Kundu}. 

We consider a system of $N$ identical bosons interacting through a two-body potential $V(\vec{r}_{ij})=V(\vec{r}_i-\vec{r}_j)$ and confined in an external harmonic potential of frequency $\omega$. The time independent quantum many-body Schr\"odinger equation is given by
\begin{eqnarray}
\Big[-\frac{\hbar^2}{2m}\sum_{i=1}^{N} \nabla_{i}^{2} 
+ \sum_{i=1}^{N} V_{trap}(\vec{r}_{i})& 
+\displaystyle{\sum_{i,j>i}^{N}} V(\vec{r}_{i}-\vec{r}_{j})\nonumber\\
-E\Big]\Psi(\vec{r}_{1},\cdots,\vec{r}_{N})=0\hspace*{.1cm};
\end{eqnarray}
where $m$ is the mass of each boson, $E$ the energy of the condensate and the trapping potential $V_{trap}(\vec{r}_i) = \frac{1}{2} m \omega^2 r^2_i$. we can eliminate the centre of mass motion by using the standard Jacobi coordinates~\cite{Fabre,Ballot,MFabre} defined as
\begin{equation}
\vec{\zeta}_{i}=\sqrt{\frac{2i}{i+1}}(\vec{r}_{i+1}-
\frac{1}{i}\sum_{j=1}^{i} \vec{r}_j) \hspace*{.5cm}
 (i=1,\cdots,{\mathcal N}),
\end{equation}
The centre of mass coordinate is $\vec{R}=\frac{1}{N}\sum_{i=1}^{N} \vec{r}_{i}$. The relative motion of the bosons is given by 
\begin{eqnarray}
\Big[-\frac{\hbar^{2}}{m}\sum_{i=1}^{\mathcal N} 
\nabla_{\zeta_{i}}^{2}+V_{trap}& + &V_{int}
(\vec{\zeta}_{1},\cdots, \vec{\zeta}_{\mathcal N})\nonumber\\
-E_{R}\Big]\Psi(\vec{\zeta}_{1},\cdots, \vec{\zeta}_{\mathcal N})& = & 
0\hspace*{.1cm}, 
\end{eqnarray} 
where ${\mathcal N} = N - 1$. Here $V_{trap}$ is the effective trapping potential and $V_{int}(\vec{\zeta}_{1},\cdots, \vec{\zeta}_{\mathcal N})$ is the sum of all pair-wise interactions expressed in terms of Jacobi coordinates. $E_{R}(=E-\frac{3}{2}\hbar\omega)$ is the relative energy of the system.

Hyperspherical harmonic expansion method (HHEM) is an {\it ab-initio} many-body tool to solve the many-body Schr\"odinger equation. The hyperspherical variables are constituted by the hyperradius $r = \sqrt{\sum_{i=1}^{\mathcal N}\zeta_{i}^{2}}$ and $(3{\mathcal N}-1)$ hyperangular variables which are comprised of $2{\mathcal N}$ spherical polar angles $(\vartheta_j,\varphi_j; \ j=1,\cdots,{\mathcal N})$ associated with ${\mathcal N}$ Jacobi vectors and $({\mathcal N}-1)$ hyperangles $(\phi_2,\phi_3,\cdots,\phi_{\mathcal N})$ giving their relative lengths. The total wave function is expanded in the complete set of hyperspherical harmonic (HH) functions~\cite{Ballot}. Therefore HHEM is a complete many-body approach and includes all possible correlations. However there are serious difficulties for a large number of particles. The calculation of potential matrix elements of all pairwise potentials becomes a formidable task and the convergence rate of the hyperspherical harmonic expansion becomes extremely slow in the large particle limit, due to rapidly increasing degeneracy of the basis. For these reasons HHEM can be used for the three-body systems only and it is not suitable for the description of the experimental BEC containing a few thousand to a few million particles. 
However for the achievement of a stable BEC in the laboratory, the atomic cloud must be extremely dilute physically, so as to preclude three-body collisions, which lead to molecule formation and consequent depletion~\cite{Dalfovo}. Hence the inter-particle separation must be very large compared to the range of the effective potential (which is $|a_s|$). Thus $n|a_s|^3 \ll 1$, where $n$ is the number density $\sim N/(a_{ho})^3$, with $a_{ho}$ being the oscillator length of the trap. In the original JILA experiment, $a_s/a_{ho}=0.00433$ and this condition is well satisfied by $N$ up to $\sim 10^6$. As a consequence three and higher body correlations are not relevant in the condensate wave function $\Psi$. Now as only two-body interactions are present, $\Psi$ can be decomposed into Faddeev components. Since only two-body correlations are relevant, the Faddeev component corresponding to the $(ij)$-interacting pair is a function of $\vec{r}_{ij}$ and $r$ only and can be expanded in the sub-set of HH, called the potential harmonics (PH) sub-set, which is sufficient for the expansion of $V(\vec{r}_{ij})$, as a function of hyperspherical variables~\cite{Tapan,Das}. This leads to a dramatic simplification: for any $N$, the active degrees of freedom is effectively reduced to only four, {\it viz.}, $\vec{r}_{ij}$ and $r$ for {\it each of the $N(N-1)/2$ Faddeev components} (the remaining irrelevant degrees of freedom in the dilute condensate being frozen). Since $\Psi$ is decomposed into all interacting pair Faddeev components, {\it all two-body correlations} are included. The emerging picture for a given Faddeev component is that when two particles interact, the rest of the particles in the condensate behave simply as inert spectators. The fact that {\it only a few  degrees of freedom are relevant} is consistent with the collective motion of the condensate as a single quantum entity.\\

Thus we decompose the total wave function $\Psi$ into two-body Faddeev components for all interacting pairs as
\begin{equation}
\Psi=\sum_{i,j>i}^{N}\phi_{ij}(\vec{r}_{ij},r)\hspace*{.1cm}\cdot
\end{equation}
As we discussed earlier, due to the presence of two-body correlations only, $\phi_{ij}$ is a function of interacting-pair separation ($\vec{r}_{ij}$) and the global hyperradius $r$ only. 
$\phi_{ij}$ is symmetric under $P_{ij}$ for bosons and satisfy the Faddeev equation
\begin{equation}
\left[T+V_{trap}-E_R\right]\phi_{ij}
=-V(\vec{r}_{ij})\sum_{k,l>k}^{N}\phi_{kl},
\end{equation} 
where $T = -\frac{\hbar^2}{m} \displaystyle{\sum_{i=1}^{\mathcal N}} \nabla_{\zeta_{i}}^{2}$ is the total kinetic energy. Applying the operator  $\sum_{i,j>i}$ on both sides of Eq.~(8), we get back the original Schr\"odinger equation. In this approach, we assume that when ($ij$) pair interacts, the rest of the bosons are inert spectators. Thus the total hyperangular momentum 
quantum number as also the orbital angular momentum of the whole system is contributed by the interacting pair only. Next  we expand $\phi_{ij}$ in the PH  subset 
\begin{equation}
\phi_{ij}(\vec{r}_{ij},r)
=r^{-(\frac{3{\mathcal N}-1}{2})}\sum_{K}{\mathcal P}_{2K+l}^{lm}
(\Omega_{\mathcal N}^{ij})u_{K}^{l}(r) \hspace*{.1cm}\cdot
\end{equation}
$\Omega_{\mathcal N}^{ij}$ denotes the full set of hyperangles in the $3{\mathcal N}$-dimensional 
space for the $(ij)$-partition, in which the $(ij)$-pair interact and 
${\mathcal P}_{2K+l}^{lm}(\Omega_{\mathcal N}^{ij})$ is a member of the PH basis. It 
has an analytic expression~\cite{Fabre} given by 
\begin{equation}
{\mathcal P}_{2K+l}^{l,m} (\Omega_{\mathcal N}^{(ij)}) =
Y_{lm}(\omega_{ij})\hspace*{.05cm} 
^{({\mathcal N})}P_{2K+l}^{l,0}(\phi) {\mathcal Y}_{0}(D-3);\hspace*{.2cm}D=3{\mathcal N},
\end{equation}
${\mathcal Y}_{0}(D-3)$ is the HH of order zero in 
the $(3{\mathcal N}-3)$ dimensional space spanned by $\{\vec{\zeta}_{1},\cdots,
\vec{\zeta}_{{\mathcal N}-1}\}$ Jacobi vectors; $\phi$ is the hyperangle given by
$r_{ij}$ = $r\hspace*{0.1cm} \cos\phi$. For the remaining $({\mathcal N}-1)$
 non-interacting bosons we define a hyperradius as
\begin{eqnarray}
 \rho_{ij}& = &\sqrt{\sum_{k=1}^{{\mathcal N}-1}\zeta_{k}^{2}}\nonumber\\
          &= & r \sin\phi \hspace*{.01 cm}\cdot
\end{eqnarray}
so that $r^2=r_{ij}^2+\rho_{ij}^2$. The relevant set of $(3{\mathcal N}-1)$ quantum numbers of HH is now reduced to {\it only} three and the remaining ones vanish 
\begin{eqnarray}
l_{1} = l_{2} =\cdots=l_{{\mathcal N}-1}=0,   & \\
m_{1} = m_{2}=\cdots=m_{{\mathcal N}-1}=0,  &   \\
n_{2} = n_{3}=\cdots=n_{{\mathcal N}-1} = 0, & 
\end{eqnarray}
and for the interacting pair $l_{\mathcal N} = l$, $m_{\mathcal N} = m$ and  $n_{\mathcal N} = K$.
Thus the $3{\mathcal N}$ dimensional Schr\"odinger equation reduces effectively
to a four dimensional equation with the relevant set of quantum 
numbers: principal quantum number $n$, orbital angular momentum quantum number $l$, azimuthal quantum number $m$ and grand orbital quantum number $2K+l$
for any $N$.
Substituting Eq.~(9) into Eq.~(8) and projecting on a particular PH, a set of 
coupled differential equation (CDE) for the partial wave $u_{K}^{l}(r)$
is obtained
\begin{equation}
\begin{array}{cl}
&\Big[-\frac{\hbar^{2}}{m} \frac{d^{2}}{dr^{2}} +
V_{trap}(r) + \frac{\hbar^{2}}{mr^{2}}
\{ {\cal L}({\cal L}+1) \\
&+ 4K(K+\alpha+\beta+1)\}-E_R\Big]U_{Kl}(r)\\
+&\displaystyle{\sum_{K^{\prime}}}f_{Kl}V_{KK^{\prime}}(r)
f_{K^{\prime}l}
U_{K^{\prime}l}(r) = 0
\hspace*{.1cm},
\end{array}
\end{equation}\\
where ${\mathcal L}=l+\frac{3N-6}{2}$, $U_{Kl}=f_{Kl}u_{K}^{l}(r)$, 
$\alpha=\frac{3N-8}{2}$ and $\beta=l+1/2$. 
$f_{Kl}$ is a constant and represents the overlap of the PH for
interacting partition with the sum of PHs corresponding  to all 
partitions~\cite{MFabre}.
The potential matrix element $V_{KK^{\prime}}(r)$ is given by
\begin{equation}
V_{KK^{\prime}}(r) =  
\int P_{2K+l}^{lm^*}(\Omega_{\mathcal N}^{ij}) 
V\left(r_{ij}\right)
P_{2K^{\prime}+1}^{lm}(\Omega_{\mathcal N}^{ij}) d\Omega_{\mathcal N}^{ij} 
\hspace*{.1cm}\cdot
\end{equation}\\ 
\subsection{Introduction of a short-range correlation function}

As the two-body interaction is represented by a contact interaction, whose strength is given by the $s$-wave scattering length $a_s$ only, the mean-field GP equation completely disregards the detailed structure of the inter-atomic  potential. The sign of $a_s$ determines nature of the interaction: a positive (negative) value of $a_s$ represents the repulsive (attractive) interaction. But in realistic inter-atomic interactions, such as the van der Waals potential, there is always an attractive $-\frac{C_6}{r_{ij}^6}$ type tail part at large separations and a strong repulsion at short separations~\cite{Pethick}. Depending on the nature of these two parts, $a_s$ can either be positive or negative. In an earlier many-body calculation \cite{tkd2008}, we observed an appreciable effect of shape-dependence of the inter-atomic interaction, for larger values of $N$, even for the dilute condensate. Moreover in the mean-field GP equation for an attractive condensate, the Hamiltonian is unbound from below and an approximate solution is obtained only in the metastable region. So in our present many-body calculation, we use a realistic inter-atomic potential, {\it viz.}, the van der Waals (vdW) potential with a hard core~\cite{Pethick}.\\

The strong short-range repulsion of the realistic potential produces a short-range correlation in the $(ij)$-interacting pair Faddeev component $\phi_{ij}$, which forbids the interacting pair to come too close. To include the effect of this strong repulsion, we introduce an additional short-range correlation function (SRCF) $\eta(r_{ij})$ in the PH expansion basis for $\phi_{ij}$. The SRCF is obtained as the zero-energy solution of the interacting pair Schr\"odinger equation
\begin{equation}
-\frac{\hbar^2}{m}\frac{1}{r_{ij}^2}\frac{d}{dr_{ij}}\left(r_{ij}^2
\frac{d\eta(r_{ij})}{dr_{ij}}\right)+V(r_{ij})\eta(r_{ij})=0
\hspace*{.1cm}\cdot
\end{equation} 
The asymptotic form of $\eta(r_{ij})$ quickly attains $C(1-a_s/r_{ij})$, from which $a_s$ is calculated~\cite{Pethick}. The hard-core radius of the vdW potential is adjusted to give the appropriate value of $a_s$. This ensures that the short-range repulsion is correctly accounted for. The inclusion of SRCF has the same effect as the Jastrow function in other many-body approaches. The zero-energy two-body wave function $\eta(r_{ij})$ is a good 
representation of the short range behavior of $\phi_{ij}$, as in the 
experimental BEC the energy of the interacting pair is negligible 
compared with the depth of the inter-atomic potential. Since $\eta(r_{ij})$ has the correct short-separation behavior of the 
$(ij)$-interacting pair, the $r_{ij} \rightarrow 0$ behavior of $\phi_{ij}$ is correctly reproduced~\cite{Kundu}. As a result the rate of convergence of the PH expansion is dramatically enhanced. 
\\

With the inclusion of SRCF, we replace Eq.~(9) by 
\begin{equation}
\phi_{ij}(\vec{r}_{ij},r)
=r^{-(\frac{3{\mathcal N}-1}{2})}\sum_{K}{\mathcal P}_{2K+l}^{lm}
(\Omega_{\mathcal N}^{ij})u_{K}^{l}(r) \eta(r_{ij}),
\end{equation}
and the correlated PH (CPH) basis function is given by
\begin{equation}
[{\mathcal P}_{2K+l}^{l,m} (\Omega_{\mathcal N}^{(ij)})]_{correlated} =
 {\mathcal P}_{2K+l}^{l,m} (\Omega_{\mathcal N}^{(ij)}) \eta(r_{ij}).
\end{equation}
The potential matrix element $V_{KK^{\prime}}(r)$ in the CPH basis is now given by
\begin{equation}
\begin{array}{cl}
&V_{KK^{\prime}}(r) =(h_{K}^{\alpha\beta} h_{K^{\prime}}^
{\alpha\beta})^{-\frac{1}{2}}\times \\
&\int_{-1}^{+1} \{P_{K}^{\alpha\beta}(z) 
V\left(r\sqrt{\frac{1+z}{2}}\right)
P_{K^{\prime}}^{\alpha \beta}(z)\eta\left(r\sqrt{\frac{1+z}{2}}\right)
W_{l}(z)\} dz. 
\end{array}
\end{equation}
Here $P_{K}^{\alpha\beta}(z)$ is the Jacobi polynomial, and its 
norm and 
weight function are $h_{K}^{\alpha\beta}$ and $W_{l}(z)$   
respectively~\cite{Abramowitz}.

Note that the inclusion of $\eta(r_{ij})$ makes the CPH basis 
non-orthogonal. One may surely use the standard procedure for handling 
non-orthogonal basis. However in the present calculation we have 
checked that $\eta(r_{ij})$ differs from a constant value only 
in a very narrow interval near the origin, in the BEC length scale $a_{ho}$ (which is much larger than the inter-atomic interaction length scale). As a result, the overlap matrix becomes a constant matrix for relevant values of $r$ (which is $\sim \sqrt{3N} \ a_{ho}$). The effect of the constant matrix is taken by a suitable choice of the asymptotic constant $C$~\cite{Sof}.


\section{Result}
\subsection{Choice of two-body potential and calculation of many-body effective potential}
As mentioned in the last section, the inter-atomic potential has been chosen as the van der Waals 
potential with a hard core of radius $r_{c}$, viz., 
$V(r_{ij}) = \infty$ for $r_{ij} \le r_{c}$ and $= -\frac{C_6}{r_{ij}^6}$
for $r_{ij} > r_{c}$. $C_6$ is known for a 
specific atom and in the limit of $C_6 \rightarrow 0$, the potential 
becomes a hard sphere and the cutoff radius exactly coincides with the
$s$-wave scattering length $a_{s}$. By utilizing the Feshbach resonance one can effectively tune
the scattering length $a_s$. In our choice of two-body potential we tune $r_c$ to reproduce the experimental 
scattering length. As we decrease $r_c$, $a_{s}$ decreases and at a 
particular critical value of $r_c$  it passes through an infinite discontinuity, going from $- \infty$ to $\infty$~\cite{Kundu}.
For our present calculation we choose Rb atoms with $C_6$ = $6.4898 \times 10^{-4}$ o.u.~\cite{Pethick}.
With this $V(r_{ij})$ we solve the zero-energy two-body Schr\"odinger 
equation and tune $r_c$ to obtain the $a_s$ correctly.
We choose $r_c$ such that it corresponds to the zero node in the zero-energy two-body 
wave function for $a_s < 0$ and one node for $a_s > 0$~\cite{Kundu, Pethick}. With these set of parameters  
we solve the coupled differential equation by
hyperspherical adiabatic approximation (HAA)~\cite{Coelho}. In HAA, 
we assume that the hyperradial motion is slow compared to the 
hyperangular motion and the potential matrix together with the 
hypercentrifugal repulsion is diagonalized for a fixed value of $r$. 
Thus the effective potential for the hyperradial motion is obtained 
as a parametric function of $r$. We choose the lowest eigen potential 
$\omega_0(r)$ [corresponding eigen column vector being $\chi_{K0}(r)$] as the effective potential in which the condensate moves
collectively. We solve the adiabatically separated hyperradial equation 
in the extreme adiabatic approximation (EAA)
\begin{equation}
\left[-\frac{\hbar^{2}}{m}\frac{d^{2}}{dr^{2}}+\omega_{0}(r)-E_{R}
\right]\zeta_{0}(r)=0\hspace*{.1cm},
\end{equation}
subject to approximate boundary condition and obtain the hyper-radial wave function $\zeta_0(r)$.
For our numerical calculation we fix $l=0$ and truncate the CPH basis 
to a maximum value $K=K_{max}$ requiring proper convergence. Finally the many-body condensate wave function $\Psi$ can be constructed in terms of $\zeta_0(r)$ and $\chi_{K0}(r)$.

The accuracy of HAA has been tested against exact results for various nuclear and atomic systems. Since exact HHE calculations are possible only for the three-body systems, the tests were done for trinuclei, two-electron atoms and exotic three-body Coulombic systems. The accuracy was found to be better than 1$\%$ even for the infinite-range Coulomb potential~\cite{HAAaccuracy}. In our calculation the van der Waals potential has shorter range and hence HAA is expected to be better~\cite{Coelho}. Moreover the confining harmonic potential is smooth and for this potential alone, the hyperradial equation is completely decoupled. Since the trapping potential has a dominant effect, this also improves the accuracy of HAA for application to BEC. For BEC with $N \leq 20$, for which exact DMC results were available, the CPHEM together with HAA were found to be very close to the exact DMC calculations~\cite{Das}. Moreover, results of our previous BEC calculations using HAA for quite large $N$, agree well with earlier calculations and experiments~\cite{Sudip1,Pankaj1,Pankaj2,Anindya1,Anindya2,Tapan,Das,Kundu,tkd2008,Sof,Anasua,Anindya3}.  Thus we can safely use HAA in our calculation and this reduces the numerical complications to a great extent. In addition, the HAA produces an effective hyperradial potential, in which the collective motion takes place.\\

For the attractive condensate, 
the many-body effective potential strongly differs from the effective 
mean-field potential. In GP, the choice of a contact $\delta$-interaction 
in the two-body potential gives rise to the pathological singularity
in the effective potential~\cite{Tapan,Kundu}. Thus the study of post-collapse scenario of 
the attractive condensate is beyond the scope of the GP theory. 
Whereas the presence of a short-range hard core in the van der Waals 
interaction does not only remove the singularity, it gives the 
realistic scenario and can describe the formation of atomic cluster 
after the collapse.

\subsection{Information entropy of one-body density}

For repulsive BEC, we consider $^{87}$Rb atoms in the external spherical harmonic trap of trap frequency 77.87 Hz. The cut off radius $r_c$ in the van der Waals potential is adjusted to get the scattering length $a_s=0.00433$ o.u. $ = 100$ a$_0$ (a$_0$ is the Bohr radius) which corresponds to the JILA experiment~\cite{Anderson}. For attractive BEC, our choice is $^{85}$Rb atoms in the JILA trap with $a_s=-1.832 \times 10^{-4}$ o.u..
From the condensate wave function $\Psi$, we calculate the one-body density $R_1(\vec{r}_k)$~\cite{Anindya3} as
\begin{equation}
 R_1(\vec{r}_{k}) = \int_{\tau^{\prime}} |\Psi|^{2} d\tau^{\prime}\hspace*{.1 cm},
\end{equation}
The one-body density basically measures the probability density of finding a particle at a distance $\vec{r}_k$ from the centre of mass of the condensate. The integral over the hypervolume $\tau^{\prime}$ excludes the variable $\vec{r}_k$ and $d\tau^{\prime}$ is given by
\begin{equation}
d\tau^{\prime} = r^{{\prime}3{\mathcal N}-4} \cos^2\phi \sin^{3{\mathcal N}-7}\phi dr^{\prime} d{\phi} d\omega_{ij} d\Omega_{{\mathcal N}-2}\hspace*{.1cm},
\end{equation}
where $r^{\prime}$ is obtained from $r^2=r^{\prime 2}+ 2 r_{k}^2$. 
After a lengthy but straight forward calculation we arrive at a close form given by
\begin{eqnarray}
R_{1}(\vec{r}_{k})=\sqrt{2}  \int_{0}^{\infty} \int_{-1}^{1} 
2^{\alpha} \left[\frac{1}{\pi^{3/2}} \frac{\Gamma\left((D-3)/2\right)}
{\Gamma\left((D-6)/2\right)}\right] \left[\zeta_{0}(r')\right]^{2}
\nonumber
\\
\sum_{KK'} \chi_{K0}(r') \chi_{K'0}(r') 
(f_{Kl}f_{K'l})^{-1}(h_{K}^{\alpha \beta} h_{K'}^{\alpha \beta})^{-1/2}  
P_{K}^{\alpha\beta}(z) 
\nonumber
\\
P_{K'}^{\alpha\beta}(z)  r'^{D-4} 
\sqrt{\frac{1+z}{2}} \left(\sqrt{\frac{1-z}{2}}\right)^{D-8}
\nonumber
\\
\left(\sqrt{r'^{2}+2r_{k}^{2}}\right)^{-(D-1)} dr' dz,~~~~~~~~~~~~
\end{eqnarray}
where $D = 3N-3$ and $h^{\alpha,\beta}_k$ is the norm of the Jacobi polynomial. The integral is computed by numerical computation using a 32-bit Gaussian quadrature. $R_1(\vec{r}_k)$ basically contains all the information of the one body density correlation. In Fig. 1(a) we plot the one body density distribution as a function of $\vec{r}_k$ (in o.u.) for several number of bososns $N=$50,500 and 5000. For small particle number the density is sharper and with increase in particle number the density distribution gradually pushed out as the effective repulsion increases. Next taking the Fourier transformation of $R_{1}(\vec{r}_{k})$, we obtain the one-body momentum distribution $\Phi_{1}(\vec{p}_{k})$ and plot in Fig.~1(b) for the same set of particle numbers as in Fig.~1(a). We see the expected reciprocal behavior between the position and momentum space wave function is accordance with the Heisenberg's uncertainty principle. For very small $N$ ($ \sim 100$), as the net effective repulsion is small, the condensate wave function is close to Gaussian in both the spaces. However, with increase in $N$, when the condensate wave function in the coordinate space spreads out due to increase in effective interaction $Na_s$, the corresponding momentum space wave function squeezes accordingly.
\begin{figure}
  \begin{center}
    \begin{tabular}{cc}
      \resizebox{80mm}{!}{\includegraphics[angle=0]{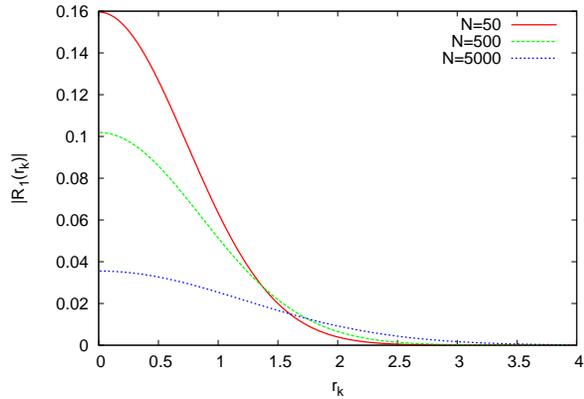}} &\\
         (a)  &\\
      \resizebox{80mm}{!}{\includegraphics[angle=0]{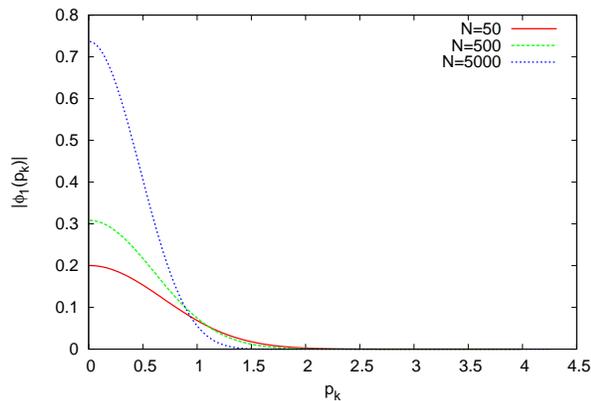}} & \\
         (b)  &\\
    \end{tabular}
  \end{center}
\caption{(Color online) Plot of one body density as a function of $r_k$ (in o.u.) and the corresponding one-body momentum distribution as a function of $p_k$ (in o.u.) for different $N$.}
\end{figure}
Next to compare our many-body results with the mean-field we numerically solve the GP equation of the form
\begin{equation}
\Big[-\frac{\hbar^2}{2m}\nabla^2
+\frac{1}{2}m\omega^2r^2
+g|\psi(\vec{r})|^2 \Big]\psi(\vec{r}) = \mu \psi(\vec{r})
\end{equation} 
for the same set of particles as chosen in the many-body calculation. Here $g=\frac{4 \pi \hbar^2 a_s {\mathcal N}}{m}$ is the mean-field interaction, $m$ being the mass of the particle. The corresponding one-body density both in the coordinate and momentum spaces are presented in Fig. 2(a) and Fig.~2(b) respectively. The expected reciprocal behavior is also seen.
\begin{figure}
  \begin{center}
    \begin{tabular}{cc}
      \resizebox{80mm}{!}{\includegraphics[angle=0]{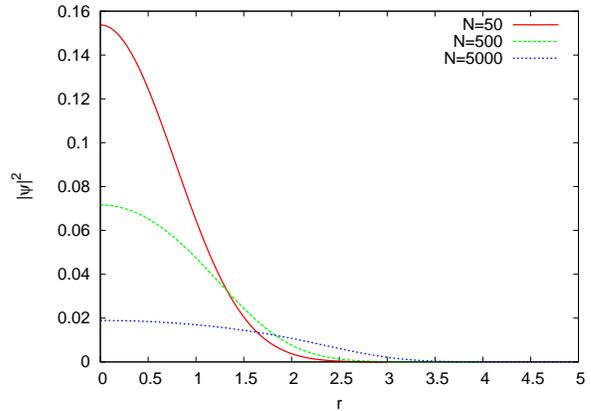}} &\\
         (a)  &\\
      \resizebox{80mm}{!}{\includegraphics[angle=0]{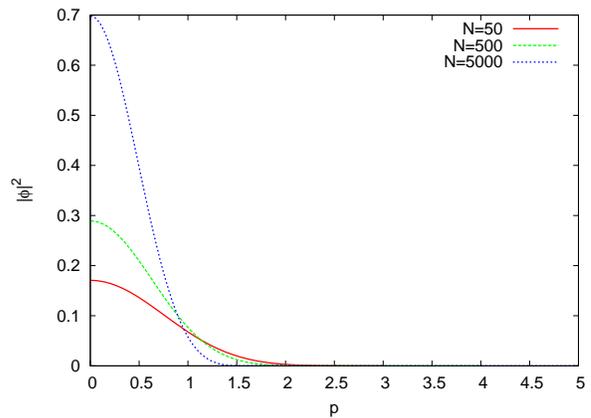}} & \\
         (b)  &\\
    \end{tabular}
  \end{center}
\caption{(Color online) Plot of condensate density $|\psi|^2$ as a function of $r$ (in o.u.) [panel (a)] and the corresponding momentum distributions $|\phi|^2$ as a function of $p$ (in o.u.) [panel (b)] obtained from mean-field GP equation for different $N$.}
\end{figure}

Next we calculate the Shannon information entropy in the coordinate space using Eq.~(1) and the same in momentum space using Eq.~(2). In Table~1 we present the values of $S_{r}$ and $S_{p}$ and the total entropy $S = S_r + S_p$ versus the number of particle $N$ for $^{87}$Rb condensate. For comparison we also calculate the same for the mean-field and present in Table 1.
\begin{center}
\begin{table}[!h]
\caption{One body entropy in coordinate space $S_{r}$, in momentum space $S_{p}$ and total one body entropy $S$ for some typical $N$ for repulsive BEC. All entropy values are given in o.u.. }
\vskip 5pt
\begin{center}
\begin{tabular} {|l|l|l|l|l|l|l|}
\hline
$N$ & \multicolumn{3}{|c|}{GP Theory}
& \multicolumn{3}{|c|}{CPHEM} \\ \cline{2-7}
     & $S_{r}$ &  $S_{p}$  & $S$      & $S_{r}$ & $S_{p}$ & $S$ \\ \hline
100  & 3.369   &  3.087    & 6.456    & 3.431   & 3.004   & 6.4343\\ \hline
500  & 3.834   &  2.630    & 6.465    & 3.782   & 2.653   & 6.435 \\ \hline
1000 & 4.102   &  2.398	   & 6.494    & 4.049 	& 2.389   & 6.437 \\ \hline		
3000 & 4.601   &  1.968    & 6.569    & 4.544   & 1.921   & 6.465\\ \hline
5000 & 4.858   &  1.750    & 6.601    & 4.804   & 1.674   & 6.478  \\ \hline
7000 & 5.032   &  1.620    & 6.627    & 4.987   & 1.495   & 6.483 \\ \hline
\end{tabular}
\end{center}
\end{table}
\end{center}
For small $N$, the net effective interaction $Na_s$ is very small compared to the trap energy ($\sim \hbar \omega$) and the total entropy is very close to the lower bound of the EUR. However with increase in particle number $N$, $S$ smoothly increases. With increase in net repulsive interaction $Na_s$, the coordinate space wave function delocalizes, consequently the position space entropy $S_r$ gradually increases. It signifies that the associated uncertainty in the coordinate space increases. The corresponding momentum space wave function is localized, the associated entropy $S_{p}$ gradually decreases with increase in $Na_s$. It indicates that the associated uncertainty in momentum space decreases. Although we observe same trend both in the many-body and mean-field results, however quantitative disagreement occurs. Next to address the universal relation for Shannon information entropy, we plot it as a function of $\ln N$ in Fig.~3 where our CPHEM result is presented as the red smooth curve and the green dashed curve corresponds to the mean-field GP result. It has been proposed in an earlier calculation~\cite{Massen} that for a system of fermions, the universal relation is maintained and has the form
\begin{equation}
S = a + b \ln N
\end{equation}
In the mean-field results (green dashed curve in Fig.~3), we retrive the straight line in the plot of $S$ vs $\ln N$ and the calculated values of $a=6.059$ and $b=0.065$. In the many-body calculation we fail to retrieve the linear relation (Eq.~(26)) for the entire particle range. However in the large particle limit we observe the straight increase in the total entropy as a function of $\ln N$. It shows that in the small and finite particle limit the many-body calculation exhibits finite-size effect, whereas in the large particle limit the mean-field results are close to the many-body results for repulsive condensates, as expected.  
\begin{figure}[hbpt]
\vspace{-10pt}
\centerline{
\hspace{-3.3mm}
\rotatebox{0}{\epsfxsize=8cm\epsfbox{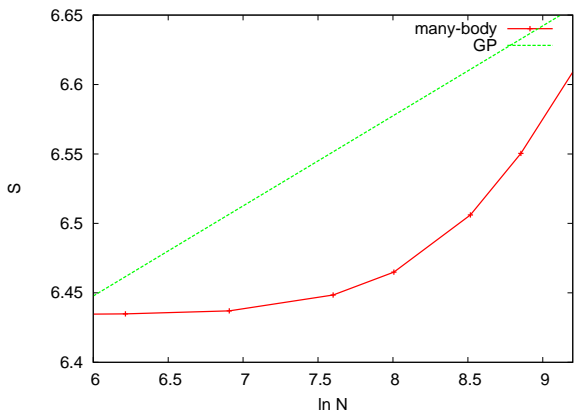}}}

\caption{(Color online) Plot of the one body entropy $S$ (in o.u.) as a function of $\ln N$. The many-body result is plotted as the red smooth curve and the green dashed curve corresponds to the mean-field GP result.}
\end{figure}

As pointed out earlier, for repulsive bosons, there is no upper limit of the number of atoms in the trap and the condensate is always stable for any number of atoms. However in the case of an attractive BEC, the condensate tends to increase its density in the centre of the trap in order to lower its interaction energy with increase in the atom number. This tendency is balanced by the zero point kinetic energy which can stabilize the system. However for large number of bosons, the central density becomes too high and the kinetic energy can not balance it anymore. The condensate thus collapses beyond a critical number $N_{cr}$ and the corresponding stability factor is defined as $k_{cr} = \frac{N_{cr}|a_s|}{a_{ho}}$. In Fig.~4 we plot $S_r$ and $S_p$ as a function of $Na_s$, where the scattering length $a_s$ is gradually tuned from large repulsive (when the condensate is absolutely stable) to attractive (when the condensate is metastable for smaller $N|a_s|$ and finally collapses at the criticality). The inverse behavior between the position and momentum space entropies is the consequence of the EUR. At $Na_s=0$, we reach the noninteracting limit when the system behaves as an ideal Bose gas (IBG). For IBG in a harmonic trap, both the position and momentum space wave functions become perfectly gaussian. Hence, the Shannon information entropy in position space $S_r$ and that in momentum space $S_p$ become equal to each other and the minimum uncertainty limit is obtained. We have checked the numerical values of $S_r$ and $S_p$. They are $S_r = S_p = 3.217$; the total entropy $S = 6.434$, which is the lower bound of EUR.
\begin{figure}[hbpt]
\vspace{-10pt}
\centerline{
\hspace{-3.3mm}
\rotatebox{0}{\epsfxsize=8cm\epsfbox{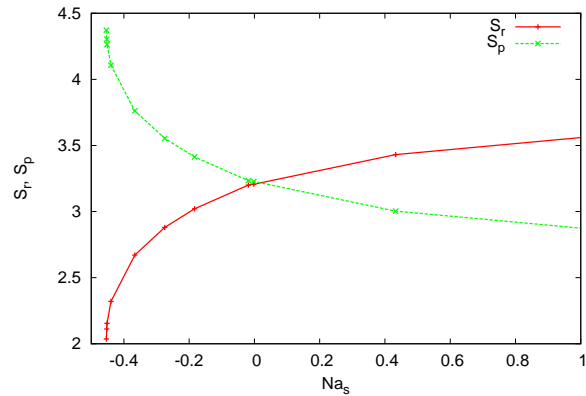}}}

\caption{(color online) Plot of the one body entropy in position space $S_r$ (in o.u.) and in momentum space $S_p$ (in o.u.)  against effective interaction $Na_s$ in o.u..}
\end{figure}
However for attractive BEC, $a_s$ is negative and the condensate is metastable as described earlier. For our present calculation we keep $a_s=-1.832 \times 10^{-4}$ o.u. which is chosen from the controlled collapse experiment~\cite{Roberts,Cornish} and go on increasing $N$ to make more and more attractive condensate. Thus the condensate density in the coordinate space contracts near the centre of the trap, whereas the corresponding momentum space wave function spreads out as expected. It signifies that the associated uncertainty in position space decreases and that in momentum space increases. Thus with increase in $N$, $S_r$ starts to fall and $S_p$ start to increase. At the critical point, the metastable region is no longer able to bound the condensate due to very strong attractive interaction energy. Hence, the condensate wave function in coordinate space squeezes to a $\delta$ function and the corresponding momentum space wave function is completely delocalised. This is reflected in a very sharp fall in $S_r$ and very steep increase in $S_p$ [Fig.~4]. Thus, at the point of collapse $S_r$ and $S_p$ diverge in opposite directions. This point of divergence is used to calculate the stability factor which we found to be $k_{cr}^{many-body}=0.457$ and is in very close agreement with the experimental value of $k_{cr}^{expt}=0.459 \pm 0.012 \pm 0.054$~\cite{Roberts}. In the mean-field GP equation stability factor is determined in the following way. When $N<N_{cr}$, the condensate is metastable and the energy functional has a local minimum~\cite{Dalfovo}. When $N$ increases, the depth of local minimum decreases and exactly at $N=N_{cr}$, the minimum vanishes and the GP equation has no solution. The calculated stability factor is $k_{cr}^{GP}=0.575$.
Thus our present many-body calculation not only calculates the stability factor accurately, but also manifests the collapse of the attractive BEC as the simultaneous divergence in $S_r$ and $S_p$.

\subsection{Information entropy of the two-body density} 

The two-body Shannon information entropies in position and momentum spaces are defined as~\cite{Guevara1}
\begin{equation}
 S_{\Gamma} = - \int R_2(r_{ij}) \ln R_2(r_{ij}) dr_{ij}\hspace*{.1cm},
\end{equation}
and 
\begin{equation}
 S_P = -\int \Phi_2(p_{ij}) \ln \Phi_2(p_{ij}) dp_{ij}\hspace*{.1cm}
\end{equation}
where $R_2(r_{ij}) $ is the two-body density distribution in the coordinate space and the corresponding pair density in momentum space $ \Phi_2(p_{ij}) $ is obtained by taking Fourier transformation of $R_2(r_{ij})$. The pair-distribution function determines the probability of finding $(ij)$th pair at a relative separation $r_{ij}$ in the coordinate space and likewise for the momentum space also. We calculate it as follows:
\begin{equation}
R_2(r_{ij}) = \int_{\tau^{\prime\prime}} |\Psi|^2 d\tau^{\prime\prime}
\end{equation}
where $\Psi$ is the many-body condensate wave function.
As before $\tau^{\prime\prime}$ excludes integration over $r_{ij}$. After a lengthy calculation we obtain closed analytic form of $R_{2}(r_{ij})$~\cite{Anindya3} as
\begin{eqnarray}
R_{2}(r_{ij})=\sqrt{2}\int_{-1}^{1}
\left(\frac{1-z}{2}\right)^{\alpha}\left(\zeta_{0}
\left(r_{ij}\sqrt{\frac{2}{1+z}}\right)\right)^{2}
\nonumber
\\
\sum_{KK'}\left(\frac{h_{K}^{\alpha\beta}}{2^{\alpha}}\right)^{-1/2}
\left(\frac{h_{K'}^{\alpha\beta}}{2^{\alpha}}\right)^{-1/2}\left(f_{Kl}f_{K'l}\right)^{-1}
\nonumber
\\
\chi_{K0}(r)\chi_{K'0}(r)
P_{K}^{\alpha\beta}(z)P_{K'}^{\alpha\beta}(z) dz.~~~~~
\end{eqnarray}
Taking Fourier transformation of $R_2(r_{ij})$  we get $\Phi_{2}(p_{ij})$ which gives the pair-density in momentum space. Now utilizing these in Eq.~(27) and Eq.~(28) we calculate $S_{\Gamma}$ and $S_{P}$ for various particle number keeping $a_s$ fixed at $0.00433$ o.u. for repulsive BEC and at $-1.832 \times 10^{-4}$ o.u. for attractive BEC. We again observe the reciprocal behavior in $S_{\Gamma}$ and $S_{P}$. $S_{\Gamma}$ gradually increases with increase in $Na_s$ and the corresponding $S_P$ gradually decreases. We again observe the diverging behavior in both $S_{\Gamma}$ and $S_{P}$ near the point of critical instability [Fig.~5]. Now it would be interesting to compare the two-body information entropies with those of one-bdoy. We observe overall similar behavior between the one-body and two-body quantities, however quantitative disagreement persists. In Fig.~4, we see that the rate of increase in $S_r$ is not completely balanced by the decrease in $S_p$, which makes an overall slow increase in the total entropy $S$. However in Fig.~5 we observe that although $S_{\Gamma}$ steeply increases with $Na_s$ as in one-body, still the rate of decrease of $S_P$ is very slow. The inter-atomic interaction with a short range hard core repulsion and long range attractive tail plays an important role here. Specially for the attractive condensate, when the atoms try to form clusters, the strong short range repulsion comes into play.
\begin{figure}[hbpt]
\vspace{-10pt}
\centerline{
\hspace{-3.3mm}
\rotatebox{0}{\epsfxsize=8cm\epsfbox{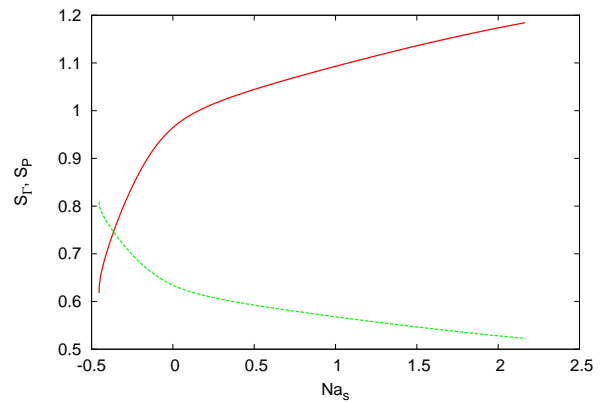}}}

\caption{(color online) Plot of pair-entropy in position space $S_{\Gamma}$ (in o.u.)(red smooth curve) and in momentum space $S_P$ (in o.u.) (green dashed curve) against the effective interaction $Na_s$ (in o.u.).}
\end{figure}
\subsection{Correlation in dilute BEC}
\begin{figure}[hbpt]
\vspace{-10pt}
\centerline{
\hspace{-3.3mm}
\rotatebox{0}{\epsfxsize=8cm\epsfbox{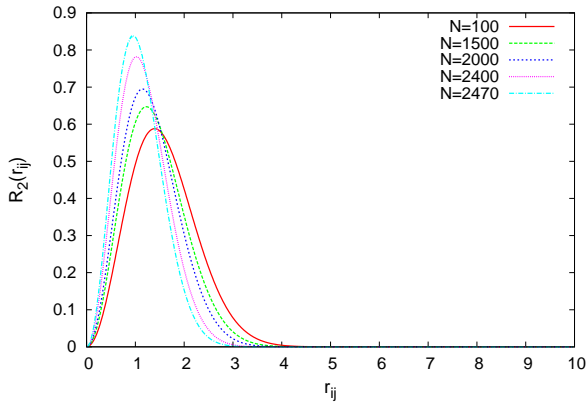}}}

\caption{(color online) . Plot of the pair distribution functions $R_2(r_{ij})$ as a function of $r_{ij}$ (in o.u.) for different $N$ for attractive BEC.}
\end{figure}
As pointed out earlier for attractive BEC, both the inter-atomic correlation and realistic inter-atomic interaction play important roles. Due to the attractive interaction, even in the weakly interacting gas, the effect of correlation becomes important and we expect to get new physics in the study of correlation properties. The mean-field Gross-Pitaevskii (GP) equation with a contact $\delta$-interaction is adequate for the description of weakly interacting repulsive Bose gas. However our correlated basis function is more rigorous and will provide a realistic picture for the study of correlation properties, especially in attractive condensates. To characterize the effect of correlations, in Fig.~6 we plot the pair-distribution function $R_2(r_{ij})$ for the attractive interaction with various particle number. Pair-correlation vanishes as $r_{ij} \rightarrow 0$, when the atoms try to form cluster due to strong inter-atomic correlation, but the  strong short range repulsion try to separate them. This realistic picture is observed due to the use of realistic inter-atomic interaction. Again due to the presence of an external trapping potential $R_2(r_{ij})$ does not extend beyond the size of the condensate. Thus $R_2(r_{ij})$ is peaked at some intermediate value of $r_{ij}$. Thus our results are different from the earlier findings of Lieb-Linger (LL) model which treats one-dimensional {\it uniform Bose gas} and particles interact via a $\delta$-function repulsive potential~\cite{Lieb, EHLieb}. For weak interactions (when the number of particles are much less than the critical number), the correlation length is large which indicates weak correlation. However for particle numbers close to the critical point, the condensate becomes strongly correlated, pair-correlation length sharply decreases and the correlation function becomes sharply localized. It leads to the possibility of formation of atomic clusters due to large two-body correlations. Next we calculate the healing length $\xi$, which is considered as the most relevant quantity to quantify the correlation of such highly correlated BEC near the critical point of collapse. It basically measures the minimum distance over which the order parameter can heal~\cite{Pethick, Dalfovo}. We calculate $\xi$ by balancing the quantum pressure and the interaction energy of the condensate and plot it in Fig.~7. $\xi$ steeply decreases with $N$ when the number of atoms is very close to the critical number.   
\begin{figure}[hbpt]
\vspace{-10pt}
\centerline{
\hspace{-3.3mm}
\rotatebox{0}{\epsfxsize=8cm\epsfbox{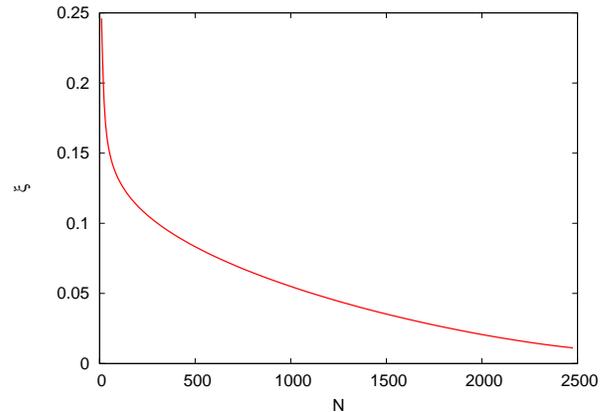}}}

\caption{(color online) Plot of healing length $\xi$ (in o.u.) with $N$ for attractive BEC.}
\end{figure}

Next we calculate another useful quantity - the correlation length $L$. We define $L$ as the half-width of correlation function and plot it in Fig.~8. The smooth decrease in $L$ with the particle number $N$ again confirms the presence of stronger correlations with increase of $N$ in the attractive BEC.  
\begin{figure}[hbpt]
\vspace{-10pt}
\centerline{
\hspace{-3.3mm}
\rotatebox{0}{\epsfxsize=8cm\epsfbox{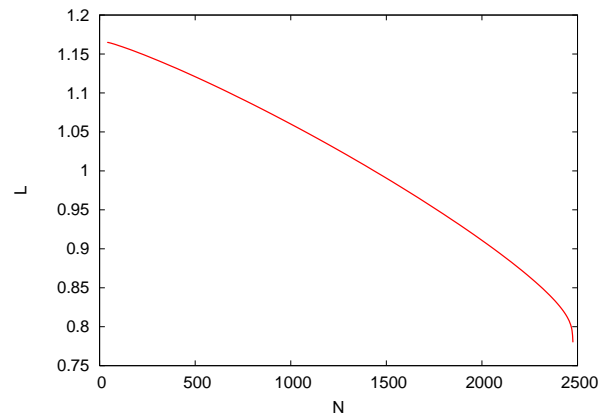}}}

\caption{(color online) Plot of correlation length $L$ (in o.u.) with $N$ for attractive BEC.}
\end{figure}

\begin{figure}[hbpt]
\vspace{-10pt}
\centerline{
\hspace{-3.3mm}
\rotatebox{270}{\epsfxsize=6cm\epsfbox{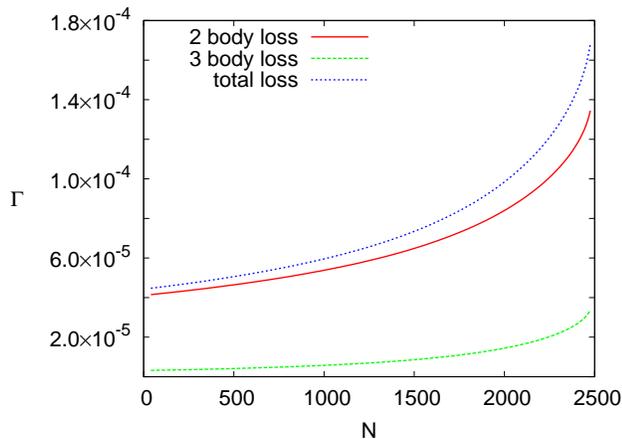}}}

\caption{(color online) Plot of the loss rate $\Gamma$ (in atoms/sec.) due to two-body and three-body collisions for various $N$ for attractive BEC.}
\end{figure}

As the attractive condensate becomes highly correlated near the critical point it is relevant to study its stability by calculating the decay rates due to two-body and three-body collisions. The loss rate due to two-body dipolar collisions and three-body recombinations are given by 
\begin{eqnarray}
 \Gamma &= &\Gamma_{two} + \Gamma_{three} \nonumber \\
 &= &K_2 \int d\tau |\psi|^{4} + K_3 \int d\tau |\psi|^{6}\hspace*{.1cm}
\end{eqnarray}
where $K_2$ is the two-body dipolar loss rate coefficient and has the value $(1.87 \pm 0.95 \pm 0.19) \times 10^{-14}$ cm$^3$/sec. The three-body recombination loss rate coefficient $K_3 = (4.24^{+0.70}_{-0.29} \pm 0.85) \times 10^{-24}$ cm$^6$/sec~\cite{Wieman}. $\psi$ is the condensate wave function in coordinate space and can be calculated as described in sec IIIB. As for an  attractive condensate, there is a rapid increase in condensate density with increase in number of atoms, we may expect a large loss rate near the critical point. It is also important to observe the separate contributions coming from the two-body dipolar loss rate and three-body recombination. In Fig.~9 we plot the two-body dipolar loss rate $\Gamma_{two}$ and the loss rate due to three-body recombination $\Gamma_{three}$. For small particle numbers, as the condensate density is not too high, the three-body loss rate is almost negligible compared to the two-body loss rate. On the other hand, near the critical point when the condensate is highly correlated, we observe a steep increase in the two-body loss rate. The contribution coming from the three-body recombination is still insignificant compared to the sharp change in the dipolar loss rate. Thus the total loss rate of the attractive condensate is dominated by the two-body loss rate only. This indicates that for the present choices of scattering length of the  controlled collapse experiment~\cite{Roberts,Cornish}, the condensate is sufficiently dilute. Fig.~9 also justifies and establishes that in sufficiently dilute limit the condensate is affected only by two-body correlations. This also justifies {\it a posteriori} our use of two-body correlated basis function (CPH). However near the Feshbach resonance, where the condensate is strongly correlated the higher-body correlations may significantly contribute and that would be the issue of future research. This is quite obvious from our present study that the effect of three-body recombination is more insignificant for repulsive BEC in the dilute condition. In the repulsive condensate, as the atoms repel each other, the central density is lower and it is less correlated than the attractive condensate. The rate-coefficient $K_3$ is significantly smaller for repulsive condensate. While for $^{85}$Rb condensate (attractive) $K_3$ is of the order of $10^{-24}$ cm$^6$/sec, it is of the order of $10^{-30}$ cm$^6$/sec for $^{87}$Rb condensate (repulsive). This clearly shows the unimportance of the three-body recombination in the context of repulsive BEC. 

\section{Conclusion}

We have reported the results of our numerical calculation on the Shannon information entropy of trapped interacting bosons and study their several characteristic features. We employ two-body correlated basis  function and realistic van der Waals interaction. This is the first such rigorous calculation where we observe the interplay between the position and momentum space information entropies. Due to presence of an external trap, the system is inhomogeneous and gives some additional features. We also numerically establish the lower bound of BBM inequality which is believed to be a stronger version of the Heisenberg uncertainty principle. The observation of curvature in the total entropy $S$ also differs from the earlier mean-field results. Our many-body results show that the finite size effect is prominent for small and finite number of atoms. We also calculate the point of critical instability of attractive BEC. When the number of atoms in the trap is very close to the critical number, the condensate becomes highly correlated and the mean-field GP equation is not enough to describe such a correlated system. Our calculated stability factor is in very close agreement with the experimental results. We observe the point of instability as the point where the information entropies in the conjugate spaces diverge simultaneously. Thus in our present calculation we not only calculate the instability accurately but at the same time we establish the point in a stronger way providing more information of the BEC as a single quantum entity. We also calculate the two-body Shannon information entropies corresponding to the two-body densities in position and momentum spaces and discuss their similarity and differences with the one-body quantities. Next we specially discuss the correlation in dilute BEC by introducing several correlation measures like healing length $\xi$, correlation length $L$. We also calculate the two-body and three-body loss rate for the attractive condensate. Very sharp increase in the decay rates signifies that near the point of criticality the condensate becomes strongly correlated and there will be a possibility of formation of atomic cluster. All these observations signify quantitatively the presence of inter-atomic correlations, even away from the critical point in an attractive BEC, which is beyond the scope of studies of mean-field theory. \\

This work has been supported by Department of Atomic Energy (DAE, India). SKH acknowledges a senior research fellowship from the Council of Scientific and Industrial Research (CSIR), India.

\end{document}